\documentclass[aps,prl,twocolumn,superscriptaddress]{revtex4-1}

\usepackage[utf8]{inputenc}
\usepackage{graphicx}
\usepackage{color}
\usepackage[tight]{units}
\usepackage{amsmath}
\usepackage{hyperref}
\newcommand{\overbar}[1]{\mkern 1.5mu\overline{\mkern-1.5mu#1\mkern-1.5mu}\mkern 1.5mu}
\begin{document}
	
	\title{Observation of an excitonic Mott transition through ultrafast core-\textit{cum}-conduction photoemission spectroscopy}
	\author{Maciej Dendzik}
	\email{dendzik@kth.se}
	\affiliation{Fritz Haber Institute of the Max Planck Society, Faradayweg 4-6, 14915 Berlin, Germany}
	\affiliation{Department of Applied Physics, KTH Royal Institute of Technology, Electrum 229, SE-16440, Stockholm, Kista, Sweden}
	\author{R. Patrick Xian}
	\affiliation{Fritz Haber Institute of the Max Planck Society, Faradayweg 4-6, 14915 Berlin, Germany}
	\author{Enrico Perfetto}
	\affiliation{CNR-ISM, Division of Ultrafast Processes in Materials (FLASHit), Area della Ricerca di Roma 1, Via Salaria Km 29.3, I-00016 Monterotondo Scalo, Italy}
	\affiliation{Department of Physics, Tor Vergata University of Rome, Via della Ricerca Scienti 1, 00133 Rome, Italy}
	\author{Davide Sangalli}
	\affiliation{CNR-ISM, Division of Ultrafast Processes in Materials (FLASHit), Area della Ricerca di Roma 1, Via Salaria Km 29.3, I-00016 Monterotondo Scalo, Italy}
	\affiliation{Department of Physics, University of Milan, via Celoria 16, I-20133 Milan, Italy}
	\author{Dmytro Kutnyakhov}
	\affiliation{DESY Photon Science, Notkestr. 85, 22607 Hamburg, Germany}
	\author{Shuo Dong}
	\author{Samuel Beaulieu}
	\author{Tommaso Pincelli}
	\affiliation{Fritz Haber Institute of the Max Planck Society, Faradayweg 4-6, 14915 Berlin, Germany}
	\author{Federico Pressacco}
	\affiliation{Center for Free-Electron Laser Science CFEL, Hamburg University, Luruper Chausee 149, 22761 Hamburg, Germany}
	\author{Davide Curcio}
	\affiliation{Department of Physics and Astronomy, Interdisciplinary Nanoscience Center (iNANO), Aarhus University, Ny Munkegade 120, 8000 Aarhus C, Denmark}
	\author{Steinn Ymir Agustsson}
	\affiliation{Institute of Physics, Johannes Gutenberg University Mainz, D-55128 Mainz, Germany}
	\author{Michael Heber}
	\affiliation{DESY Photon Science, Notkestr. 85, 22607 Hamburg, Germany}
	\author{Jasper Hauer}
	\affiliation{Fritz Haber Institute of the Max Planck Society, Faradayweg 4-6, 14915 Berlin, Germany}
	\author{Wilfried Wurth}
	\affiliation{DESY Photon Science, Notkestr. 85, 22607 Hamburg, Germany}
	\affiliation{Center for Free-Electron Laser Science CFEL, Hamburg University, Luruper Chausee 149, 22761 Hamburg, Germany}
	\author{Günter Brenner}
	\affiliation{DESY Photon Science, Notkestr. 85, 22607 Hamburg, Germany}
	\author{Yves Acremann}
	\affiliation{Department of Physics, Laboratory for Solid State Physics, ETH Zurich,  Otto-Stern-Weg 1, 8093 Zurich, Switzerland}
	\author{Philip Hofmann}
	\affiliation{Department of Physics and Astronomy, Interdisciplinary Nanoscience Center (iNANO), Aarhus University, Ny Munkegade 120, 8000 Aarhus C, Denmark}
	\author{Martin Wolf}
	\affiliation{Fritz Haber Institute of the Max Planck Society, Faradayweg 4-6, 14915 Berlin, Germany}
	\author{Andrea Marini}
	\affiliation{CNR-ISM, Division of Ultrafast Processes in Materials (FLASHit), Area della Ricerca di Roma 1, Via Salaria Km 29.3, I-00016 Monterotondo Scalo, Italy}
	\author{Gianluca Stefanucci}
	\affiliation{Department of Physics, Tor Vergata University of Rome, Via della Ricerca Scienti 1, 00133 Rome, Italy}
	\affiliation{INFN, Sezione di Roma Tor Vergata, Via della Ricerca Scienti 1, 00133 Rome, Italy}
	\author{Laurenz Rettig}
	\author{Ralph Ernstorfer}
	\email{ernstorfer@fhi-berlin.mpg.de}
	\affiliation{Fritz Haber Institute of the Max Planck Society, Faradayweg 4-6, 14915 Berlin, Germany}
	
	\date{\today}
	\begin{abstract}
		Time-resolved soft-X-ray photoemission spectroscopy is used to simultaneously measure the ultrafast dynamics of core-level spectral functions and excited states upon excitation of excitons in WSe$_2$. We present a many-body approximation for the Green's function, which excellently describes the transient core-hole spectral function. The relative dynamics of excited-state signal and core levels reveals a delayed core-hole renormalization due to screening by excited quasi-free carriers, revealing an excitonic Mott transition. These findings establish time-resolved core-level photoelectron spectroscopy as a sensitive probe of subtle electronic many-body interactions and an ultrafast electronic phase transition.
	\end{abstract}
	\maketitle

	Optoelectronic properties of semiconductors are largely governed by two types of excitations -- excitons~\cite{Mueller18}, the bosonic quasiparticles comprised of an electron and a hole bound by Coulomb interaction, and quasi-free carriers (QFCs) of single-particle character~\cite{Steinhoff17,Guerci19}. While the interplay between excitons and QFCs has been studied experimentally with terahertz and optical spectroscopies~\cite{Kaindl03,Huber01}, these techniques are restricted to optically allowed transitions and do not provide direct information about the underlying many-body interactions. 
	In this letter, we show that detailed information about the dynamics of both excitons and QFCs can be deduced from the simultaneous measurement of the core-hole spectral function and the excited state population with ultrafast time-resolved X-ray photoelectron spectroscopy (trXPS)~\cite{Pietzsch08}. 
	We observe strong renormalization of the W~$4f$ spectral function after optical excitation of WSe$_2$ bulk crystals. 
	The transient spectral function is excellently reproduced using a many-body approximation for the Green’s function~\cite{Stefanucci13}, which accounts for the core-hole screening by photo-excited QFCs. The simultaneous measurement of the excited-state population in the conduction band (CB) reveals an $\sim$100~fs delay of the core-hole screening compared to the initial build-up of exciton population, which we ascribe to an ultrafast Mott transition from optically-prepared excitons to an uncorrelated QFC plasma.

	Static XPS has been a workhorse of surface science by driving the understanding of catalytic processes~\cite{Ertl79,Ertl80,Asahi01}, chemical states of interfaces~\cite{Biesinger11}, and functional materials~\cite{Xu09}. The measured photoelectron distribution is proportional to the core-hole spectral function and carries information about the many-body interactions such as Auger scattering, electron-phonon coupling, plasmonic excitations and local screening~\cite{Citrin77,Hofner03,Mahan00,Lizzit98}. 
	The XPS lineshape of metals is usually asymmetric and phenomenologically well-described by the Doniach-Šunjić (DS) function~\cite{Doniach70}, where the characteristic heavy tail towards higher binding energy originates from the core-hole screening by conduction electrons. For semiconductors, on the other hand, the observed lineshape is typically symmetric and can be described by a Voigt profile~\cite{Dendzik15,Dendzik17}.
	In the presence of excited carriers, a semiconductor becomes partially-metallic and one can expect a renormalization of the core-hole lineshape. This opens up the possibility of studying non-equilibrium dynamics with XPS. Recently, technological advances of femtosecond X-ray sources~\cite{Ackermann07} and photoelectron detectors~\cite{Schonhense15} enabled ultrafast trXPS experiments to be conducted. These include the observation of melting of charge-density wave states in Mott insulators~\cite{Hellmann10,Ishizaka11}, charge-transfer dynamics at semiconductor interfaces~\cite{Siefermann14} or transient surface-photovoltage control~\cite{Liu18}. Nevertheless, an accurate theoretical description of the out-of-equilibrium core-hole spectral function is still missing and applying the DS theory to the dynamic case is problematic. In this work, we generalize the DS theory to cover the case of a photo-excited semiconductor, which enables a quantitative description of the fundamental processes governing the experimentally observed core-hole spectral changes.

\begin{figure}
		\includegraphics[width=0.48\textwidth]{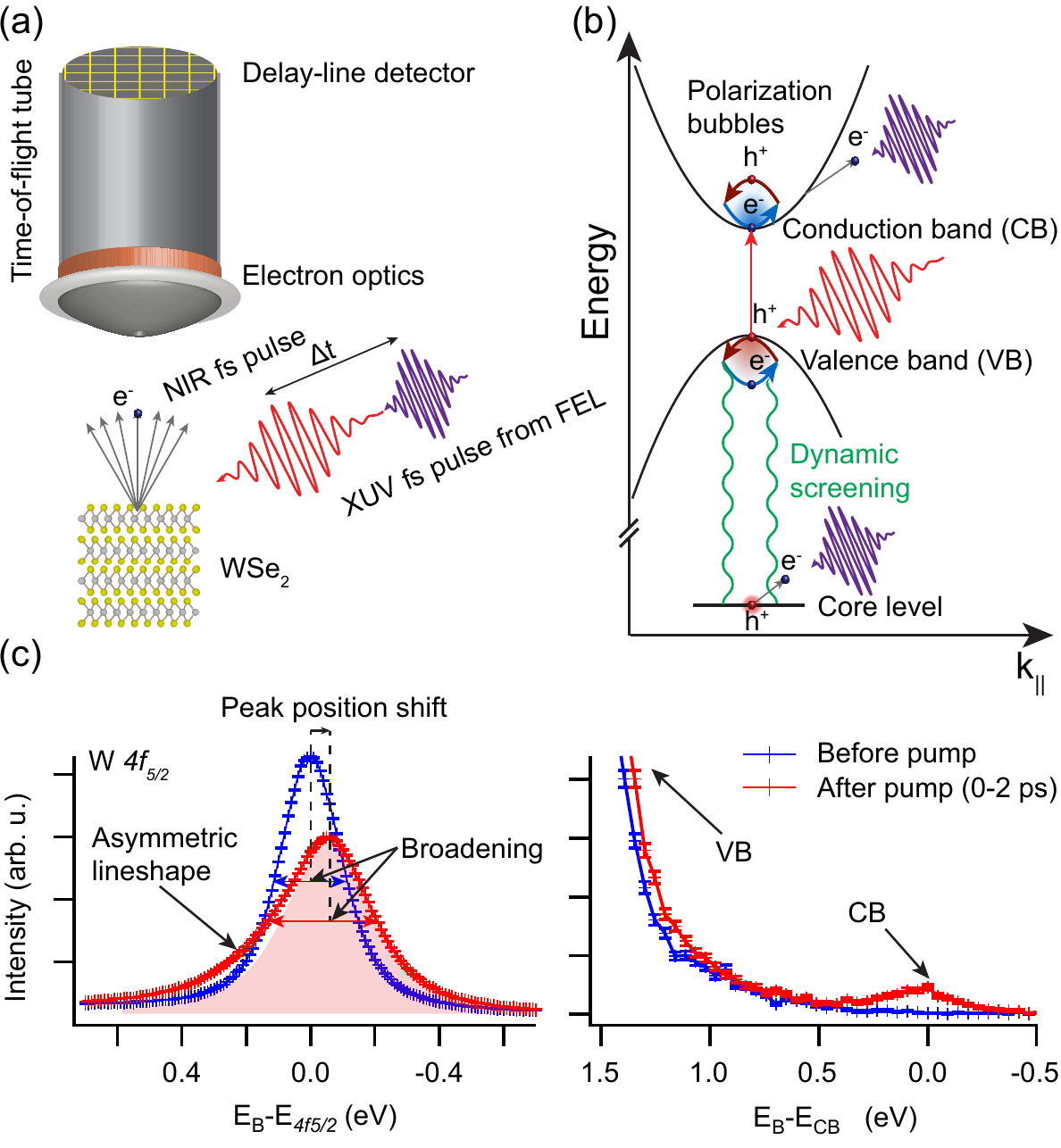}\\
		\caption{(a) Sketch of the experimental setup for pump-probe trXPS on WSe$_2$ using a momentum microscope, which enables the simultaneous analysis of core, valence and conduction electrons. A sample is excited with a near-infrared pulse (NIR) and probed with an extreme ultraviolet (XUV) pulse. (b) Energy band diagram of a photoexcited semiconductor: the pump pulse generates excited carrier populations in the valence and conduction bands. Transient conduction band population as well as the induced renormalization of the screened core-hole spectral function are detected by the probe pulse. (c) Core-\textit{cum}-conduction trXPS spectra showing W~$4f_{5/2}$ (left) and conduction band (right) regions, before (blue) and after the excitation (red). Colored arrows indicate corresponding full-width at half maximum of the spectra and the shaded area is the symmetric part of the lineshape, illustrating its asymmetry.}
		\label{fig:fig1}
	\end{figure}

	We performed core-\textit{cum}-conduction trXPS experiments of WSe$_2$ using the FLASH free-electron laser and optical pump pulses tuned to the optical A-exciton resonance at 1.6~eV at room temperature. A time-of-flight (ToF) momentum microscope was used as a photoelectron analyzer, which enabled us to simultaneously probe a $\sim$40~eV-broad spectral window including the excited states, valence band and the highest W and Se core levels ~\cite{SM,Kutnyakhov20}. Schematics of the setup and of a model conduction-core energy level diagram are shown in the Figs.~\ref{fig:fig1}(a-b). The observed time-dependent trXPS spectra of W $4f_{5/2}$ shown in Fig.~\ref{fig:fig1}(c) exhibit distinct dynamics with respect to pump-probe delay time, showing characteristic changes of photoemission peak position and width. In addition, we observed a build-up of asymmetry (skewness) which resembles the DS asymmetry. Simultaneosly, we observed a transient population of excited carriers which is responsible for the core-hole lineshape modifications.  
	
	In order to understand the origin of the observed trXPS spectral changes, we propose a theoretical model to describe the dynamical screening of the core hole due to the photo-induced valence holes and conduction electrons. We refer the reader to~\cite{SM} for details. Briefly, the core-level photoemission signal is proportional to the core-hole 
	spectral function~\cite{Hofner03}
	\begin{equation}
	A(\omega)=-\pi^{-1}\operatorname{Im}[G(\omega)],
	\label{eq:eq1}    
	\end{equation}
	with $G(\omega)$ being the core-hole Green's function:
	\begin{equation}
	G(\omega)=\frac{1}{\omega-\epsilon_c -\Sigma(\omega)+i\gamma}.
	\label{eq:eq2}    
	\end{equation}
	Here, $\epsilon_c$ is the core energy, $\Sigma(\omega)$ is the correlation self-energy due to scatterings between the core electron and conduction/valence electrons and $\gamma$ quantifies the broadening due to other decay channels, such as Auger or phonon scattering. According to the DS theory, the non-interacting lineshape is mainly renormalized by dynamical screening effects. In the diagramatic formalism, this means that the self-energy $\Sigma(\omega)$ is 
	dominated by the GW term~\cite{Hedin1971}, where $W$ is the screened interaction in the random phase approximation. In this work, we show that screening the interaction with the single polarization bubble of QFCs (see Fig.~\ref{fig:fig1}b) is enough to reproduce the core-level shift and the asymmetric lineshape. The screening due to excitons is much weaker in comparison to QFCs for transition metal dichalcogenide (TMDC) materials~\cite{Steinhoff17} and it is therefore neglected. 
	The resulting self-energy takes the form
	\begin{equation}
	\Sigma(\omega)=\lambda \log{\left(\frac{D}{\omega-\tilde{\epsilon}_c+i\gamma}\right)},
	\label{eq:eq5a}
	\end{equation}
	where $D$ is a parameter proportional to the average of conduction and valence bandwidths, while the 
	renormalized core energy reads
	\begin{equation}
	\tilde{\epsilon}_c=\epsilon_c+\lambda L\left(\frac{e^{\lambda/D}D}{\lambda}\right)
	\quad;\quad
	\lambda=\frac{m^*}{\pi}n_{QFC}\nu^2.
	\label{eq:eq5b} 
	\end{equation}
	In Eq.~(\ref{eq:eq5b}), $L(x)$ is the Lambert function, $m^*$ is the effective mass at the band edge (the average value of conduction and valence band effective masses), $n_{QFC}$ is the quasi-free carrier density and $\nu$ is the average Coulomb interaction between the core electron and the valence/conduction electrons. In the absence of QFCs, $A(\omega)$ reduces to a Lorentzian profile with width dictated by $\gamma$, while at a finite QFC density, the real and imaginary parts of the self-energy are responsible for the shift of the core energy and the asymmetric lineshape.
	
	\begin{figure*}
		\includegraphics[width=1\textwidth]{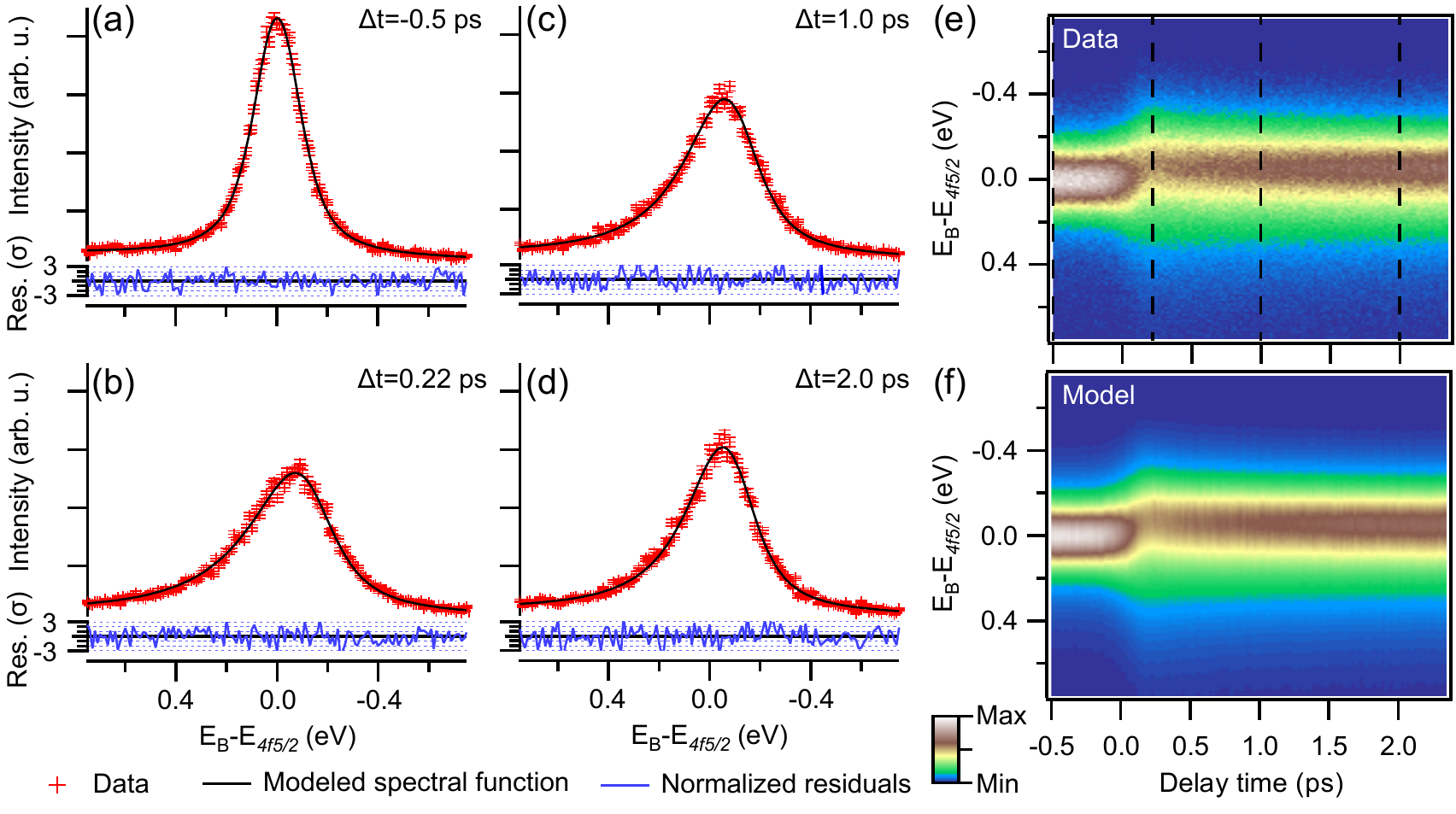}\\
		\caption{(a) Representative energy spectra at four delay times marked in (e). Red points, black and blue lines mark experimental data, fit result and normalized residuals, respectively. Normalized residuals are presented in multiples of the data's Poisson distribution standard error. (e) trXPS spectrum of W~$4f_{5/2}$ as a function of pump-probe delay time. Photoemission intensity is encoded in the false-color scale. (f) Corresponding modeled spectral function, defined by Eqs.(\ref{eq:eq1}-\ref{eq:eq5b}).}
		\label{fig:fig2}
	\end{figure*}

\begin{figure*}
		\includegraphics[width=1\textwidth]{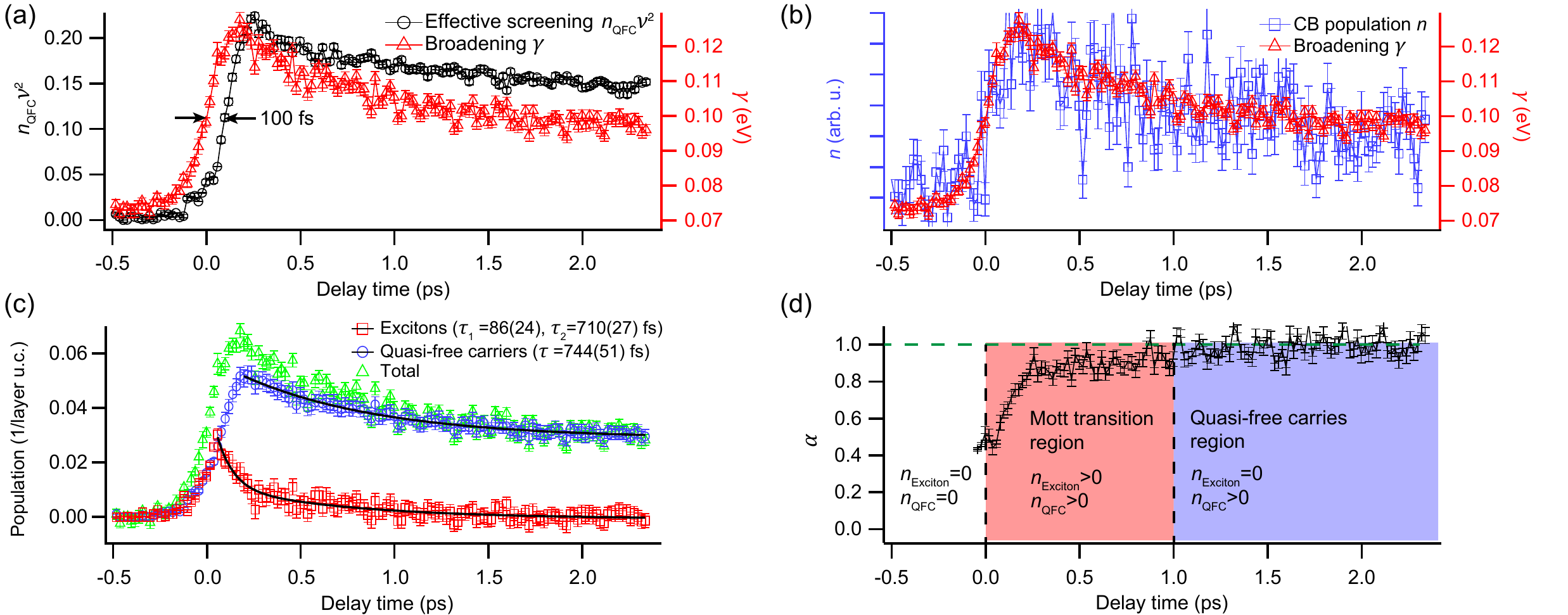}\\
		\caption{(a) Comparison of the time-dependent effective screening $n_{QFC}\nu^2$ (black circles) and broadening $\gamma$ (red triangles), obtained from fitting the theoretical model to the experimental data presented in Fig.~\ref{fig:fig2}. (b) Comparison of the photoemission signal above the Fermi level, corresponding to the CB population, (blue squares) and $\gamma$ (red triangles). (c) Decomposition of the dynamics of excitons (red squares), QFC (blue circles) and total excited carriers (green triangles). Black lines present fitted double exponential and single exponential decay for excitons and QFC, correspondingly. (d) The degree of ionization dynamics, as defined in the main text. Three different temporal regions are marked by the background color: before the excitation (white), Mott transition (red) and QFC regions (blue).}
		\label{fig:fig3}
	\end{figure*}

	The modeled spectral function (Eqs.~\ref{eq:eq1}-\ref{eq:eq5b}), convoluted with a constant Gaussian to account for the experimental energy resolution, excellently reproduces the experimental data, as presented in Fig.~\ref{fig:fig2}. This is evidenced by the featureless normalized residuals, shown for four representative time delays in Figs.~\ref{fig:fig2}(a-d). The entire time series (see Figs.~\ref{fig:fig2}(e-f)) can be reproduced by fixing the average bandwidth $D=0.8$ and effective mass $m^*=0.5$ of the material and solely fitting the broadening $\gamma(t)$ and the product $n_{QFC}(t)\nu^2$
	(the average interaction $\nu$ is independent of time) for every delay time $t$. The transient spectral function can thus be described by just two independent parameters, as $n_{QFC}\nu^2$ couples the experimentally observed peak shift and asymmetry of the lineshape and $\gamma$ describes the symmetric broadening. Interestingly, these parameters exhibit drastically different dynamics, with $n_{QFC}\nu^2$ rising ca.~100~fs later and decaying slower than $\gamma$ (see Fig.~\ref{fig:fig3}(a)). This effect is not induced by the applied model, as the same behaviour is present in model-independent quantities such as the higher moments of the photoelectron distribution~\cite{SM}. 
	
	The ToF momentum microscope allows simultaneous detection of photoelectrons over a large energy range, spanning electrons from the core levels, valence band and excited population in the CB, within a single experiment (see fig.~\ref{fig:fig1}). Therefore, it is possible to directly compare the dynamics of $\gamma$ with the build-up of the excited-state population $n$, and we find a strong correlation between the two quantities (see Fig.~\ref{fig:fig3}(b)), i.e.~the core-level broadening immediately follows the buildup of excited carriers $n$ which includes contributions from both excitons~\cite{Christiansen19} and QFCs in the CB. In contrast, the core-hole lineshape renormalization governed by the quasi-particle screening $n_{QFC}\nu^2$ shows a clear delay in buildup compared to $\gamma$ and $n$. This is consistent with the prediction that the pump energy tuned to the excitonic resonance should favor the creation of excitons~\cite{Perfetto19} up to a critical density~\cite{Steinleitner17}, and can be explained by means of an excitonic Mott transition -- the initial stage of the dynamics is dominated by excitons which subsequently break into a QFC plasma. An estimation of the excitation density per layer,  $n=7(1.4)\times10^{13}$~cm$^{-2}$~\cite{SM}, used in our experiment indeed significantly exceeds the predicted critical excitation density of approximately $3\times10^{12}$~cm$^{-2}$ ~\cite{Steinhoff17}, and is close to the density of $1.1\times10^{14}$~cm$^{-2}$ reported for experimental observation of excitonic Mott transition in single-layered WS$_2$~\cite{Chernikov15}.

	The simultaneous acquisition of both excited states population in the whole surface Brillouin zone and renormalized core-hole spectral function enables us to exclude the effect of space-charge, often observed in ultrafast photoemission experiments~\cite{Hellmann12,Oloff14}, as space-charge would not contribute to the CB population. We also exclude the influence of the inter-band $\overbar{K}-\overbar{\Sigma}$ scattering due to much faster dynamics of ca.~15~fs~\cite{Puppin18}. Moreover, we exclude surface-photovoltage observed before for WSe$_2$~\cite{Liu18} as origin of the observed renormalization. This effect can influence the peak position, but not the asymmetry of the XPS spectra. Finally, the effect of laser-assisted photoemission is minimized by the choice of s-polarization for the pump. All these observations strongly suggest the electronic excitation as origin of the lineshape renormalization.      
	
	Based on the Mott transition interpretation and the assumption that screening by excitons is negligible compared to QFCs~\cite{Steinhoff17}, we can effectively disentangle both of these populations, as presented in the Fig.~\ref{fig:fig3}(c)~\cite{SM}. The result indicates that the excitonic population reaches the critical value within the pump pulse envelope, which is then followed by a rapid decay with a lifetime of $\tau_1=86(24)$~fs. The remaining exciton population decays at much lower rate ($\tau_2=710(27)$~fs). In contrast, the QFC population continues to rise even after excitation as a result of exciton dissociation, and decays with a lifetime $\tau=744(51)$~fs. Interestingly, this decay seems to correlate with the slow component of exciton decay which is likely due to diffusion into the bulk, as excited-carrier lifetimes are typically observed on $\sim$ 100 ps time scale~\cite{Tokudomi07}. 
	
	Previous experimental studies of the excitonic Mott transition in TMDCs have observed either continuous~\cite{Chernikov15} or discontinuous behaviour~\cite{Bataller19}, while theory predicts that both of these cases can be realized depending on the interaction strength~\cite{Guerci19}. Our results shed new light into the ultrafast dynamics of the exciton-QFC transition (see Fig. ~\ref{fig:fig3}(d)). The observed degree of ionization $\alpha=n_{QFC}/n$ stays in the range of 0.5-1 for the first 1 ps after photo-excitation. For longer time delays, the excited carriers are found exclusively in the QFC state. The observation is indicative of a continuous transition with coexisting phases. It should be underlined that details of the transition will depend on the excitation density. Additional data acquired simultaneously, but with 25\% lower pump fluence indicate that both $\alpha$ and the initial decay of exciton population depend on the total excitation density, corroborating our data interpretation~\cite{SM}.
	
	\begin{figure}
		\includegraphics[width=0.48\textwidth]{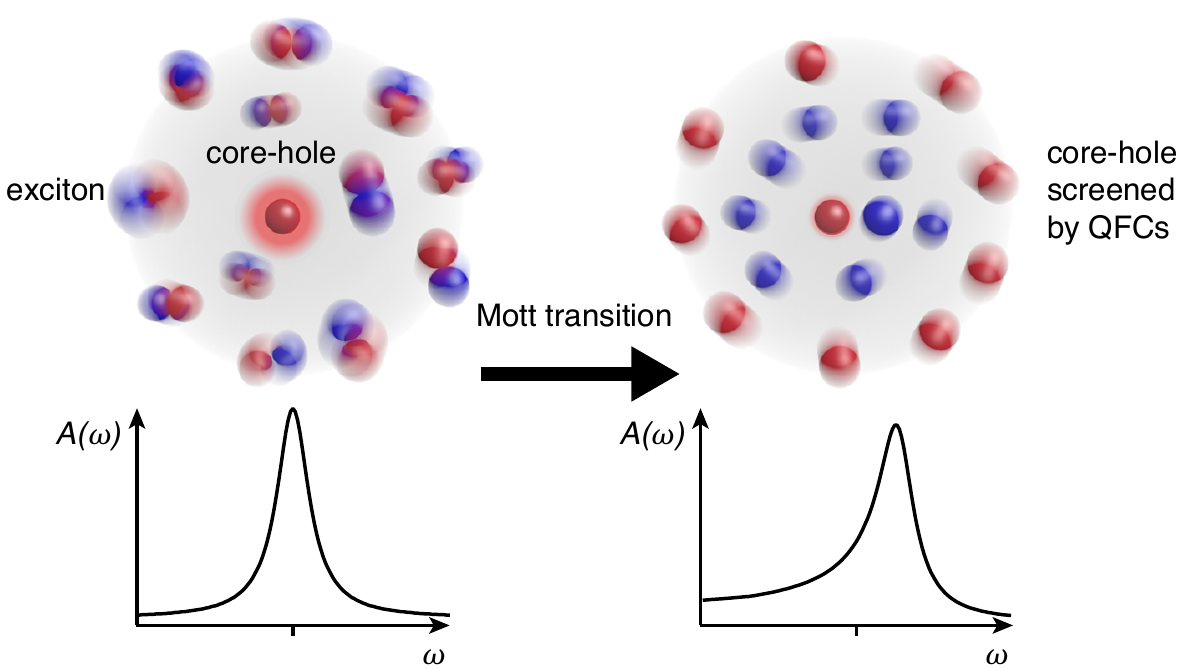}\\
		\caption{Illustration of the effect of excitons and QFCs on the core-hole lineshape. Charge-neutral electron-hole pairs only weakly screen the core holes and have a negligible effect on its spectral function (left). In contrast, single-particle-like QFCs more effectively screen the localized charge residing in the core hole resulting in a renormalization of the photoemission lineshape (right).}
		\label{fig:fig4}
	\end{figure}
	The possibility to disentangle the dynamics of excitons and QFCs by trXPS is quite surprising due to the small energy difference between these two phases, typically $\sim$50 meV, in comparison to the characteristic core-state energy scale of tens of eV. However, it was shown before that small changes in the valence band structure can have a dramatic influence on the shape of core-level spectra~\cite{Dendzik17}. Discrimination of excitons and QFCs is based on the different screening of the core-hole created during the photoemission process (see Fig~\ref{fig:fig4}). This can be understood quite intuitively -- excitons, being localized charge dipoles, are expected to interact much weaker with the suddenly created core-hole potential than delocalized QFCs. The most cogent manifestation of this effect is the $\sim100$~fs delay of the core-level peak-position shift with respect to the peak-width increase (see Fig.~\ref{fig:fig3}(a) for the dynamics of closely-related parameters). The detailed interpretation of this result relies on the proposed theoretical model which ,we believe, is general enough  to be successfully applied to further ultrafast studies of many-body states as well as electronic phase transitions. This seems to be especially appealing in combination with hard x-ray photoemission due to its larger probing depth, providing access to buried interfaces in realistic semiconducting devices.
	
	This work is dedicated to Wilfried Wurth, who passed away on May 8, 2019.
	We acknowledge support by the scientific and technical staff of FLASH and valuable discussions with Karsten Horn and Lucia Reining.
	
	This work was funded by the European Research Council (ERC) under the European Union’s Horizon 2020 research and innovation program (Grant No. ERC-2015-CoG-682843, No.~676598, No.~824143, and No.~654360), the German Research Foundation (DFG) within the Emmy Noether program (Grant No.~RE 3977/1), through the SFB/TRR 227 "Ultrafast Spin Dynamics" (projects A09 and B07), the SFB 925 "Lichtinduzierte Dynamik und Kontrolle korrelierter Quantensysteme" (project B2), and the Max Planck Society.
	F.P.~acknowledges funding from the excellence cluster “The Hamburg Centre for Ultrafast Imaging - Structure, Dynamics and Control of Matter at the Atomic Scale” of the Deutsche Forschungsgemeinschaft (DFG EXC 1074).  M.D., R.E., and L.R.~acknowledge funding from NFFA Europe (project 879). G.S., D.S., A.M., and E.P.~acknowledge funding from MIUR PRIN Grant No.20173B72NB. This work was supported by VILLUM FONDEN via the Centre of Excellence for Dirac Materials (Grant No. 11744).
	\section*{References}
\bibliographystyle{apsrev}
\bibliography{references}

\end{document}

\title{Observation of an excitonic Mott transition through ultrafast core-\textit{cum}-conduction photoemission spectroscopy: Supplementary Material}
%\title{Ultrafast dynamics of the excitonic Mott transition in WSe$_2$ studied by time-resolved XPS}
%Femtosecond x-ray photoelectron spectroscopy
\author{Maciej Dendzik}
\email{dendzik@kth.se}
\affiliation{Fritz Haber Institute of the Max Planck Society, Faradayweg 4-6, 14915 Berlin, Germany}
\affiliation{Department of Applied Physics, KTH Royal Institute of Technology, Electrum 229, SE-16440, Stockholm, Kista, Sweden}
\author{R. Patrick Xian}
\affiliation{Fritz Haber Institute of the Max Planck Society, Faradayweg 4-6, 14915 Berlin, Germany}
\author{Enrico Perfetto}
\affiliation{CNR-ISM, Division of Ultrafast Processes in Materials (FLASHit), Area della Ricerca di Roma 1, Via Salaria Km 29.3, I-00016 Monterotondo Scalo, Italy}
\affiliation{Department of Physics, Tor Vergata University of Rome, Via della Ricerca Scienti 1, 00133 Rome, Italy}
\author{Davide Sangalli}
\affiliation{CNR-ISM, Division of Ultrafast Processes in Materials (FLASHit), Area della Ricerca di Roma 1, Via Salaria Km 29.3, I-00016 Monterotondo Scalo, Italy}
\affiliation{Department of Physics, University of Milan, via Celoria 16, I-20133 Milan, Italy}
\author{Dmytro Kutnyakhov}
\affiliation{DESY Photon Science, Notkestr. 85, 22607 Hamburg, Germany}
\author{Shuo Dong}
\author{Samuel Beaulieu}
\author{Tommaso Pincelli}
\affiliation{Fritz Haber Institute of the Max Planck Society, Faradayweg 4-6, 14915 Berlin, Germany}
\author{Federico Pressacco}
\affiliation{Center for Free-Electron Laser Science CFEL, Hamburg University, Luruper Chausee 149, 22761 Hamburg, Germany}
\author{Davide Curcio}
\affiliation{Department of Physics and Astronomy, Interdisciplinary Nanoscience Center (iNANO), Aarhus University, Ny Munkegade 120, 8000 Aarhus C, Denmark}
\author{Steinn Ymir Agustsson}
\affiliation{Institute of Physics, Johannes Gutenberg University Mainz, D-55128 Mainz, Germany}
\author{Michael Heber}
\affiliation{DESY Photon Science, Notkestr. 85, 22607 Hamburg, Germany}
\author{Jasper Hauer}
\affiliation{Fritz Haber Institute of the Max Planck Society, Faradayweg 4-6, 14915 Berlin, Germany}
\author{Wilfried Wurth}
\affiliation{DESY Photon Science, Notkestr. 85, 22607 Hamburg, Germany}
\affiliation{Center for Free-Electron Laser Science CFEL, Hamburg University, Luruper Chausee 149, 22761 Hamburg, Germany}
\author{Günter Brenner}
\affiliation{DESY Photon Science, Notkestr. 85, 22607 Hamburg, Germany}
\author{Yves Acremann}
\affiliation{Department of Physics, Laboratory for Solid State Physics, ETH Zurich,  Otto-Stern-Weg 1, 8093 Zurich, Switzerland}
\author{Philip Hofmann}
\affiliation{Department of Physics and Astronomy, Interdisciplinary Nanoscience Center (iNANO), Aarhus University, Ny Munkegade 120, 8000 Aarhus C, Denmark}
\author{Martin Wolf}
\affiliation{Fritz Haber Institute of the Max Planck Society, Faradayweg 4-6, 14915 Berlin, Germany}
\author{Andrea Marini}
\affiliation{CNR-ISM, Division of Ultrafast Processes in Materials (FLASHit), Area della Ricerca di Roma 1, Via Salaria Km 29.3, I-00016 Monterotondo Scalo, Italy}
\author{Gianluca Stefanucci}
\affiliation{Department of Physics, Tor Vergata University of Rome, Via della Ricerca Scienti 1, 00133 Rome, Italy}
\affiliation{INFN, Sezione di Roma Tor Vergata, Via della Ricerca Scienti 1, 00133 Rome, Italy}
\author{Laurenz Rettig}
\author{Ralph Ernstorfer}
\affiliation{Fritz Haber Institute of the Max Planck Society, Faradayweg 4-6, 14915 Berlin, Germany}
\date{\today}
%\begin{abstract}
%Optical properties of semiconductors are strongly influenced by excitons -- the %fundamental bosonic states comprising of electron and hole bound by Coulumb %interactions. For sufficiently high excitation density, excitons are prone to Mott %dissociation into the quasi-free carrier (QFC) plasma. This could be potentially used %to transiently modify the optical properties of an opto-electronic device by simply %changing the intensity of irradiating light. In order to study the dynamics of %excitonic Mott transition we developed a general model describing the renormalization %of core-hole spectral line-shape due to the screening of excited carriers and applied %it to the WSe$_2$ time resolved x-ray photoelectron spectroscopy (trXPS) data. We %find that significant difference in screening strength of excitons and quasi-free %carriers enables to disentangle the sub-picosecond dynamics of both those species %during the Mott transition.

%Furthermore, we find that A-exciton population in WSe$_2$ decays on subpicosecond %timescale for excitation density of ca. 10$^{14}$~cm$^{-2}$.               
%\end{abstract}
\maketitle
\section{Theoretical model}
\subsection{Renormalization of core lineshape in  metals at 
equilibrium}
\label{sec1}

In this Section we put forward a minimalistic diagramatic approximation to describe the 
well known {\it Doniach-\v Sunji\'c effect}~\cite{Doniach70} (DSE), 
i.e. the screening-induced renormalization of the core level lineshape 
occurring in photoemission from metals. The DSE consists in a {\it shift}  of the core level energy
toward lower binding energy accompanied by an 
{\it asymmetric broadening}
of the core-level lineshape.
We then extend the theory to semiconductors in a quasi-stationary excited
state and provide a simple formula to fit the experimental data.

In a metal at equilibrium, the photoemission signal from a core level 
is proportional to the core-hole 
spectral function
\begin{equation}
    A(\omega)=-\pi^{-1}\operatorname{Im}[G(\omega)],
\label{eq:eq1}    
\end{equation}
with $G(\omega)$ the core-hole Green's function
\begin{equation}
    G(\omega)=\frac{1}{\omega-\epsilon_c -\Sigma(\omega)+i\gamma}.
\label{eq:eq2}    
\end{equation}
In the above equation, $\epsilon_c$ is the unscreened core-level energy, 
$\Sigma(\omega)$ is the correlation self-energy due to electron-electron 
interactions, $\gamma$ is an additional broadening due to other decay 
channels such as phonon scatterings.
According to the Doniach-\v Sunji\'c theory, the main 
renormalization of the non-interacting lineshape comes from the dynamical screening
of electron-hole pairs around the Fermi energy.
In the diagramatic language, this means that the self-energy
$\Sigma(\omega)$ is dominated by the $GW$ term~\cite{hedin},
where the screened interaction $W$ is obtained by 
summing the random phase approximation (RPA) series of electron-hole polarization bubbles.
Indeed, the $GW$ self-energy 
correctly produces (i) a {\it shift} of the core energy 
$\e_{c} \to \tilde{\e}_{c}$ toward lower binding energy
(i.e.  $\e_{c} < \tilde{\e}_{c}$), and (ii) an {\it asymmetry} in the lineshape,
characterized by a more pronounced broadening for $\omega < \tilde{\e}_{c}$~\cite{hedin,Gerlach01,Borgatti18}. 

We now show that both effects (i)-(ii) can actually be captured  by screening the interaction 
with only one polarization bubble, $W\to W^{(1)}$.
This simplification allows us to derive a simple analytical expression for $\Sigma$
in terms of few physical parameters that can be fitted with 
the experimental data.
In a two-dimensional system, the $GW^{(1)}$ self-energy of the core electron reads
\be
\Sigma(\omega)=i \int \frac{d\omega'}{2\p}
\frac{d\blq}{(2\pi)^{2}} G(\omega'+\omega)W^{(1)}(\blq,\omega')
\label{2Bself0}
\ee
with 
$W^{(1)}(\blq,\omega')=2i\th(\omega')v^{2}_{\blq}\,\Im[\chi_{0}(\blq,\omega')]$. 
To lowest order, the dynamical screening is quadratic in the repulsion 
\be
v_{\blq} = \int d\blr 
d\blr'\varphi^{*}_{c}(\blr)\varphi^{*}_{\blk}(\blr')\frac{1}{|\blr-\blr'|}
\varphi_{\blk+\blq }(\blr') \varphi_{c}(\blr) 
\ee
between a core electron with wavefunction $\varphi_{c}(\blr)$ 
and a conduction electron with wavefunction $\varphi_{\blk}(\blr)$.
The full $GW$ self-energy is obtained by replacing the Lindhard 
response function $\chi_{0}$ with the RPA one $\chi=\chi_{0}+\chi_{0}v\chi$, omitting the dependence on 
frequency and momentum. Inserting Eq.~(\ref{2Bself0}) into Eq.~(\ref{eq:eq2}), we obtain a nonlinear equation for $G$ to be solved self-consistently. 

Because our purpose is an analytic formula for the experimental fit, we have only partially dressed the Green's function. In the  first step, we have evaluated the self-energy with the bare Green's function, $G^{(0)}(\omega)=1/(\omega-\e_{c}+i\gamma)$, and found
\be
\Sigma[G^{(0)}](\omega)=2 \int \frac{d\blk}{(2\pi)^{2}}  
\frac{d\blq}{(2\pi)^{2}}  v^{2}_{\blq}  \frac{f_{\blk} (1-f_{\mathbf{k+q}})}{  
\omega+\ve_{\mathbf{k+q}}-\ve_{\blk} -\e_{c}+i\gamma },
\label{2Bself}
\ee
where $\ve_{\blk}$ is the energy of a conduction state with 
quasimomentum $\blk$ and $f_{\blk}$ is the Fermi function evaluated at $\ve_{\blk}$. In the zero-temperature limit and at low density, the integral in Eq.~(\ref{2Bself}) can be evaluated analytically, assuming a quadratic dispersion $\ve_{\blk}=|\blk|^{2}/2m^*$ and a momentum-independent {\it average repulsion} $v_{\blq} \approx \nu$:
\begin{equation}
     \Sigma[G^{(0)}](\omega)=\lambda \log{\left(\frac{D}{\omega-\epsilon_c+i\gamma}\right)}
     \label{eq:eq5a}
\end{equation}
where $D$ is a parameter proportional to the bandwidth while 
$\lambda=\frac{m^*}{\pi}n\nu^2$ is an effective interaction
depending on the density $n$ of conducting electrons. 
We have used Eq.~(\ref{eq:eq5a}) to obtain a one-shot dressed Green's function, $G^{(1)}(\omega)=1/(\omega-\tilde{\e}_{c}+i\gamma)$, where 
$\tilde{\e}_{c}$ is the solution of $\omega-\e_{c}-\Re[\Sigma[G^{(0)}](\w)]=0$, which is, explicitly,
\begin{equation}
\tilde{\epsilon}_c=\epsilon_c+\lambda 
L\left(\frac{e^{\lambda/D}D}{\lambda}\right),
\label{etilde}
\end{equation}
with $L(x)$ being the Lambert function. Subsequently, we have evaluated the self-energy using $G^{(1)}$ and inserted the result into Eq.~(\ref{eq:eq2}) to obtain the partially-dressed Green's function,
\begin{equation}
    G(\omega)=\frac{1}{\omega-\epsilon_c -\lambda 
    \log{\left(\frac{D}{\omega-\tilde{\epsilon}_c+i\gamma}\right)}+i\gamma}.
\label{Gfinal}    
\end{equation}
This simple analytic form correctly describes the shift of the 
core energy toward lower binding energy as well as the asymmetric lineshape,
% , while the frequency-dependence of $\Sigma$ in Eq.~\ref{eq:eq5a}
% is such that the resulting lineshape becomes asymmetric, 
with a longer tail for $\w < \tilde{\epsilon}_c$, which is in agreement with the Doniach-\v Sunji\'c theory.

\subsection{Renormalization of core lineshape in  semiconductors out of equilibrium}
\label{sec2}

We now extend the above analysis to semiconductors of gap $\D$ and driven out of equilibrium by an optical pulse. It is clear that if the semiconductor is in its ground state, the 
effects discussed in the previous section are negligible.
Indeed, the polarization bubble in Eq.~(\ref{2Bself}) does not induce any sizable reshaping of the core spectral function $A(\w)$ at frequencies close to $\e_{c}$, since the energy of particle-hole excitations $\ve_{\mathbf{k+q}}-\ve_{\blk} \geq \D$ cannot be arbitarily small.  

The situation is, however, completely different if the semiconductor is in an excited state with a finite electron-density $n_{c}$ in the conduction band and a finite hole-density $n_{v}=n_{c}$ in the valence band. In this case, the system can accommodate again particle-hole 
excitations with vanishing energy, giving rise to a finite
Doniach-\v Sunji\'c renormalization.

Below we assume that the system is in a {\it quasi-stationary} excited state described by two different Fermi distributions, $f_{v \blk}$ and $f_{c \blk}$ for the valence and conducting bands, respectively \cite{Perfetto16}. In particular, the functions, $f_{i \blk}$, are characterized by different Fermi energies, $\ve_{F i}$, and different temperatures. In order to make use of the analytic results derived in the previous section, we assume that $T_{c}=T_{v}=0$ and that the core electron interacts {\it separately} with electrons in the valence or conduction band, with the same average repulsion $\nu$. In this case, it is immediately seen that the non-equilibrium total self-energy $\Sigma$ takes on two additive contributions, originating from particle-hole excitations around $\ve_{Fv}$ and $\ve_{Fc}$, respectively.
For low excited densities, we can approximate the valence dispersion as $\ve_{v \blk} \approx -\D/2 -|\blk|^{2}/2m^{*}$, and the conduction dispersion as $\ve_{c\blk} \approx \D/2 +|\blk |^{2}/2m^{*}$ to obtain a self-energy identical to the one in Eq.~(\ref{eq:eq5a}) and Eq.~(\ref{etilde}), with the formal replacement $n\to n_{v}+n_{c}=2 n_c$. In general, if more bands are involved, the total self-energy is obtained by setting $n\to 2 n_{QFC}$, where $n_{QFC}$ is the density of quasi-free carriers (QFC) in the conduction bands. Note that the self-energy correctly vanishes in the semiconductor ground state, where $n_{QFC}=0$. Thus in a semiconductor, the Doniach-\v Sunji\'c renormalization can occur similarly as in a metal, provided that a finite carrier density is promoted from the valence to the conducting band.

In the experiment considered in the present work, the complex dynamics following the photo-excitation can be described by the two time-dependent parameters, $n_{QFC}(t)$ and $\gamma(t)$,
while the rest of parameters remain constant. Thus we can extract the explicit values of $n_{QFC}(t)$ and $\gamma(t)$ 
by fitting the time-dependent spectral function $A(\w,\t)$
with the experimental lineshape at each time delay $\t$.
\section{Time-resolved X-ray photoemission spectroscopy}
trXPS measurements were performed at the PG2 beamline of FLASH free-electron laser (FEL) operating with effective 5 kHz repetition rate~\cite{Kutnyakhov20}. Commercially available (HQ Graphene) single-crystal WSe$_2$ samples were cleaved \textit{in situ} and measured at room temperature. Measurements were performed with 110 eV s-polarized light, using the HEXTOF momentum microscope instrument~\cite{Kutnyakhov20}. For near-infrared (NIR) pumping, a laser system based on optical parametric chirped pulse amplification synchronized with the FEL pulses was employed. The photon energy spectrum of the pump centered around 1.6~eV was measured before and after the trXPS measurements. The pump beam-size on the sample (ca. 520$\times$114~$\mu$m$^2$) was estimated based on the pump-induced multiphoton-photoemission footprint measured using the momentum microscope. The calibration of the image was performed using a Chessy test-sample (Plano GmbH). The total absorbed fluence of 1.7(0.34)~mJ/cm$^2$ was calculated using the Fresnel equations, taking into account the geometry of the experiment and optical loses. Details of the pump fluence estimation can be found in the excitation density estimation section. Overall, the obtained energy and temporal resolutions were 130~meV and 160~fs, respectively.

Every photo-electron event recorded by the delay-line detector was stored with corresponding beam-diagnostic information, such as beam-arrival monitor (BAM) or pump photo-diode signal. The presented results show data which was filtered with respect to the pump fluence and corrected for FEL jitter (BAM correction~\cite{Schulz15}), spherical timing aberration and probe-induced space-charge~\cite{Kutnyakhov20,Xian19}. The analysis was performed using an open-source software package developed for multi-dimensional photoemission data processing~\cite{Xian19}. Additional information about data processing can be found in the data processing section.

The disentanglement of exciton and QFC populations is based on transients of two observables: $n_{QFC}\nu^2$ and $\gamma$. As the dynamics of $\gamma$ follows the CB population (see Fig.~3(b)), the total excitation density $n=n_{QFC}+n_{Ex}$ is obtained by linear scaling of the $\gamma$ signal to match the measured absorbed fluence in the first layer of the material (see the excitation density estimation section). The units of $n$ are chosen to be carriers per first-layer unit cell. The total effective screening from both QFCs and excitons is assumed to be a sum of two contributions $n\nu^{*2}=n_{QFC}\nu^2+n_{Ex}\nu_{Ex}^2$, with $\nu^{*}$ being the mean interaction strength. Under the assumption that screening due to excitons is much less effective than screening due to QFCs, i.e. $\nu_{Ex}<<\nu$, $n\nu^{*2}\approx n_{QFC}\nu^2$. Dividing $n\nu^{*2}$ by $n$ yields the transients of $\nu^{*}$ which shows a constant value for $t>1$~ps. The average value over the range [1.5,2] ps of 2.21(0.13)~eV is taken as an estimator for $\nu$. The QFC population $n_{QFC}$ is then directly obtained from $n_{QFC}\nu^2$ and the exciton population from $n_{Ex}=n-n_{QFC}$.       
\section{Model-independent analysis}
\begin{figure*}[ht]
	\includegraphics[width=1\textwidth]{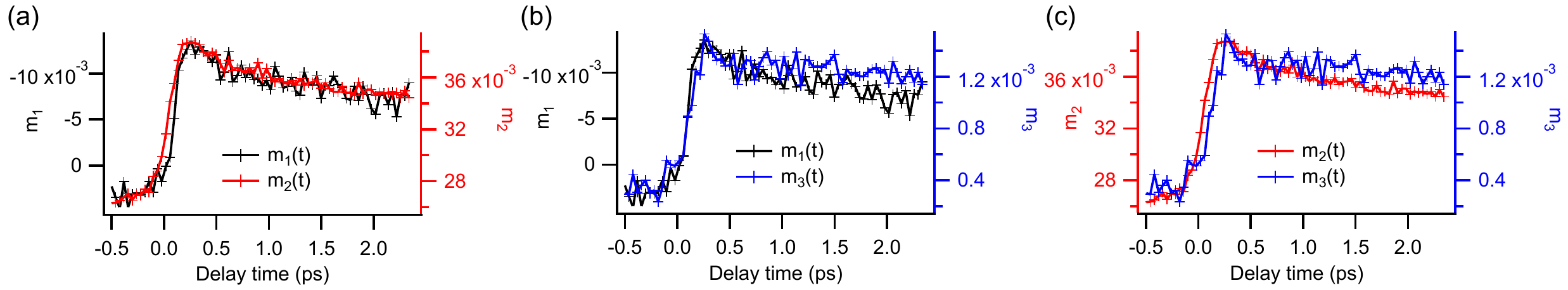}\\
	\caption{Comparison of W $4f_{5/2}$ peak statistical central moments dynamics:  $m_1$ and $m_2$-(a), $m_1$ and $m_3$-(b),  $m_2$ and $m_3$-(c). See text for details.}
	\label{fig:fig1_sm}
\end{figure*}

We investigate the influence of the proposed model on the fitting results in order to exclude any numerical artifacts or correlations of model parameters, which could affect the physical interpretation. Specifically, we calculate the model-independent, purely statistical central moments $m_n(t)$ of measured W $4f_{5/2}$ photoelectron energy distribution curves $EDC(\omega,t)$, according to the formulas,
\begin{equation}
   m_0(t)=\int_{\omega_{min}}^{\omega_{max}} I(\omega,t)d\omega,
\label{eq:eq1a_sm}
\end{equation}

\begin{equation}
   m_1(t)=m_0(t)^{-1}\int_{\omega_{min}}^{\omega_{max}} \omega I(\omega,t)d\omega,
\label{eq:eq1b_sm}
\end{equation}

\begin{equation}
   m_n(t)=m_0(t)^{-1}\int_{\omega_{min}}^{\omega_{max}} (\omega-m_1(t))^n I(\omega,t)d\omega,\; n=2,3. 
\label{eq:eq1c_sm}
\end{equation}
Here, the terms $m_{1,2,3}(t)$ correspond to the mean (peak position, with subscript 1), variance (peak width, with subscript 2) and skewness (peak asymmetry, with subscript 3), respectively. Fig.~S\ref{fig:fig1_sm} shows the comparison of time-dependent moments, calculated for the same dataset as presented in the main text. During the initial stage of the excitation, $m_1(t)$ and $m_3(t)$ are correlated (Fig.~S\ref{fig:fig1_sm}(b)) and rise slower than $m_2(t)$, which is in agreement with the modeled spectral function results. This parallel observation proves that the description of two dominant contributions affecting the lineshape renormalization, which we attributed to the excitonic Mott transition, as well as their temporal separation, is model-independent.     

\section{Excitation density estimation}
A central point of the presented work is the disentanglement of excitonic and QFC populations, which is based on accurate excitation density estimation. This section describes technical details of the absorbed fluence determination. 
\subsection{Pump beam size}
\begin{figure*}[ht]
	\includegraphics[width=1\textwidth]{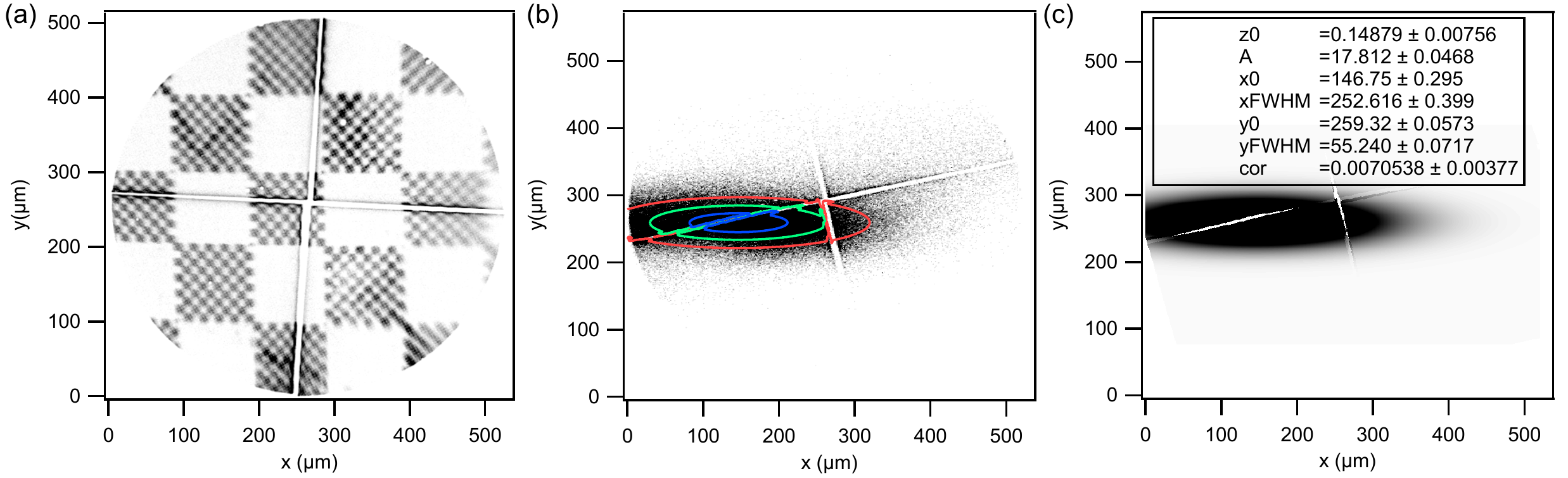}\\
	\caption{(a) Real-space image of the Chessy calibration sample acquired using the momentum microscope. (b) Multi-photon photoemission image of the pump-beam footprint. (c), 2D Gaussian fit to the image shown in (b). Table presents fitting coefficients. Constant-intensity contours of the fit in (c) are superimposed onto the data in (b).}
	\label{fig:fig2_sm}
\end{figure*}

We estimated the pump footprint on the sample based on multiphoton photoemission originating from the pump beam alone, using the real-space (PEEM) mode of the momentum microscope. Firstly, we calibrated the image scale with a test specimen (Chessy, Plano GmbH), consisting of well-defined Au squares on a Si substrate (see Fig.~S\ref{fig:fig2_sm}(a)). Secondly, we imaged the pump multiphoton photoemission distribution at the measurement position using the same lens settings, as presented in Fig.~S\ref{fig:fig2_sm}(b). The pump profile was then fitted with a 2D Gaussian function (Fig.~S\ref{fig:fig2_sm}(c)). 
It should be noted that the actual beam size is larger than the obtained profile due to the nonlinearity of the multiphoton photoemission process. The intensity of the multiphoton photoemission $I_{MP}$ signal can be expressed as,   
\begin{equation}
   I_{MP}\propto I^p,
\label{eq:eq1_sm} 
\end{equation}
where $I$ is the light intensity and $p$ is the order of the transition, unknown to us \textit{a priori}. For a Gaussian pump profile with a waist $\varpi$, one also obtains,
\begin{equation}
   \varpi=\sqrt{p}\varpi_{MP}.
\label{eq:eq2_sm} 
\end{equation}
In order to obtain the effective order of the transition under the given experimental conditions, one can vary the light intensity and track the number of photoelectrons, since
\begin{equation}
   \log(I_{MP})\propto p\log(I).
\label{eq:eq3_sm} 
\end{equation}

\begin{figure*}[ht]
	\includegraphics[width=1\textwidth]{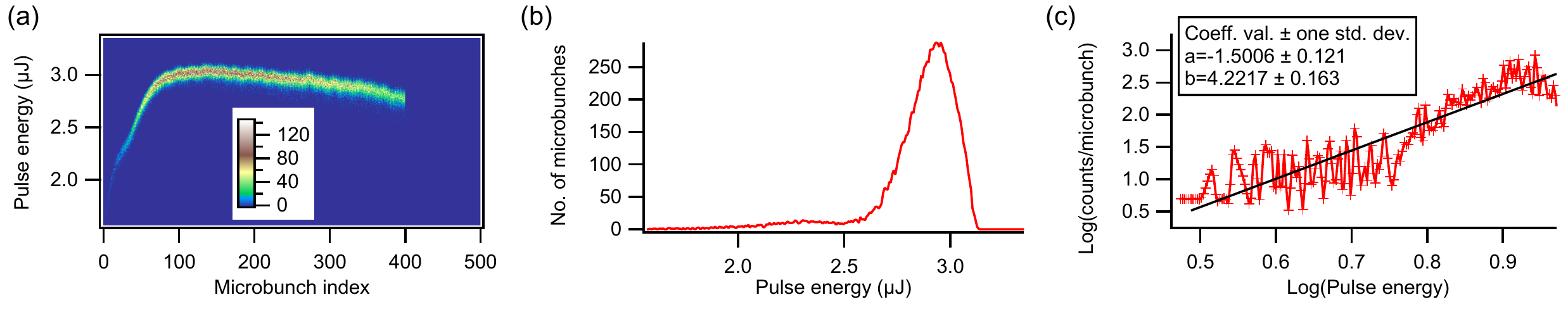}\\
	\caption{(a) 2D histogram of the acquired pump multi-photon photoemission data as a function of pulse energy and microbunch index. False-color scale represents the number of events. (b) Number of microbunches of the same pulse energy used for normalization of the count rate. (c) Log-log plot showing the dependence of the count-rate per microbunch versus pulse energy together with a linear fit and at table showing the fitting results.}
	\label{fig:fig3_sm}
\end{figure*}

We estimate the effective order of photoemission by exploiting the versatile capabilities of the experimental setup. The pump pulses from the OPA are synchronized to the microbunches from the FEL and every detected photoelectron is recorded with the corresponding pump diode readings (calibrated externally for pulse energy) and microbunch index. In Fig.~S\ref{fig:fig3_sm}(a), 2D histogram of photoelectron yield plotted as function of pulse energy and microbunch index clearly shows that the pump intensity is not constant within a macrobunch, and has a regime of rising pump intensitiy within the first 100~microbunches, before reaching a plateau for the remaining 300~macrobunches (the last 100 microbunches are unpumped). One can normalize the above histogram with the number of microbunches of the same pump intensity (Fig.~S\ref{fig:fig3_sm}(b)) to obtain the dependence between the average count rate per microbunch and the pulse energy. This procedure gives $p=4.22(0.16)$ (Fig.~S\ref{fig:fig3_sm}(c)) and a beam size (FWHM) of $519\times114~\mu$m, with estimated uncertainty of 10\%. The above result is confirmed by independent knife-edge measurement giving $500\times200~\mu$m.          
\subsection{Estimation of absorbed fluence}
\begin{figure*}[ht]
	\includegraphics[width=1\textwidth]{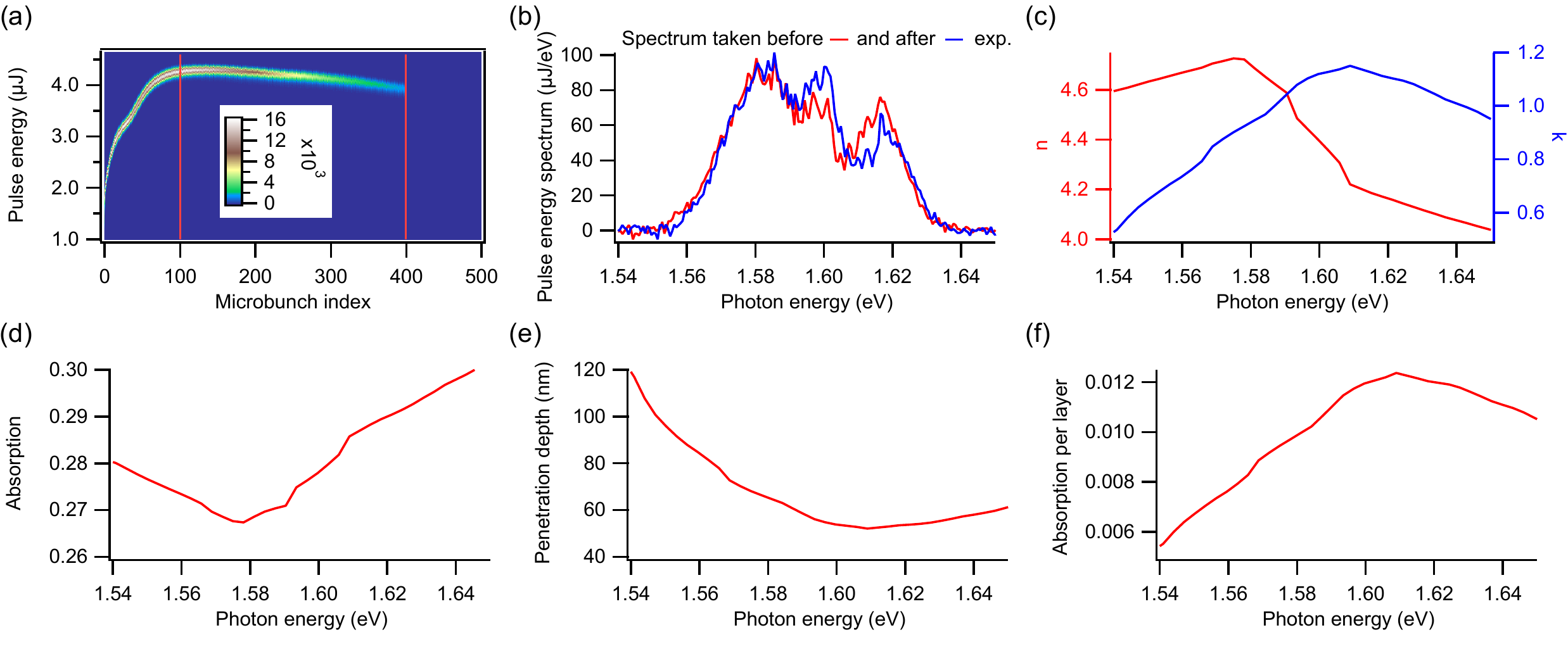}\\
	\caption{(a) 2D histogram of the W4f core-level data as a function of pulse energy and microbunch index. False-color scale represents the number of events. Red lines mark the [100,399] microbunch index range used for the data shown in the main text. (b) Photon energy spectrum of the pump beam acquired before (red) and after (blue) the trXPS experiment. (c) Complex refractive index of WSe$_2$. (d-f) Calculated absorption, penetration depth and absorption per layer, respectively.}
	\label{fig:fig4_sm}
\end{figure*}

Fig.~S\ref{fig:fig4_sm}(a) shows the pulse energy distribution for trXPS measurements of W $4f$ presented in the main text. In order to ensure a constant pump intensity, only photoelectrons with microbunch indices in the range 100-399 are included in the analysis. These photoelectrons are excited by pump pulses with an average pulse energy of 4.2(0.11)~$\mu$J. We use this value to normalize the pulse energy spectra in Fig.~S\ref{fig:fig4_sm}(b), obtained before and after the trXPS experiment. 

For $s$-polarized light at grazing incidence (AOI=68$^{\circ}$), most of the intensity is reflected. We take this into account by solving the Fresnel equations using tabulated values for the bulk WSe$_2$ complex refractive index~\cite{Beal76} $\tilde{n}(\omega)=n(\omega)+ik(\omega)$ (Fig.~S\ref{fig:fig4_sm}c) for the pump spectral range (Fig.~S\ref{fig:fig4_sm}(d)). The penetration depth $l(\omega)$ is calculated using~\cite{Dresselhaus18}
\begin{equation}
   l(\omega)=\frac{c}{2\omega k(\omega)}\sqrt{1-\frac{\sin^2{AOI}}{n(\omega)^2}},
\label{eq:eq4_sm} 
\end{equation}
resulting in the pump energy-dependent absorption, $ABS(\omega)$, within the first monolayer with a thickness of $d$=6.5~$\AA$,
\begin{equation}
   ABS(\omega)=1-\exp{[-d/l(\omega)]},
\label{eq:eq5_sm} 
\end{equation}
as presented in Fig.~S\ref{fig:fig4_sm}(e-f), respectively. The obtained absorption coefficients are multiplied with the measured pulse energy spectrum, integrated and divided by the beam size, yielding the excitation density in the first layer of $n=0.7(0.14)\times10^{14}$~cm$^{-2}$.

\subsection{Uncertainty estimation}
Estimation of the experimental uncertainty of the excitation density is summarized in Tab.~\ref{tab:tab1_sm}. Based on this analysis, we determine the maximum uncertainty of 20\%.
\begin{table}[]
\begin{tabular}{ll}
\textbf{Source of uncertainty} & \textbf{Value (\%)} \\
Beam size                      & 10                  \\
Power meter                    & 2                   \\
Absorption                     & 2                   \\
Penetration depth              & 2                   \\
Laser stability                & 3                   \\
Other systematic errors        & 1                  
\end{tabular}
\caption{\textbf{Summary of various kinds of uncertainties affecting the excitation density.}}
\label{tab:tab1_sm} 
\end{table}

\section{Data processing}
Data acquisition based on single-event detection enables corrections for many experimental artifacts. This, however, needs to be applied carefully as error-detection is not straightforward for the case of multidimensional datasets. This section describes the details of the preprocessing of the single-event data which led to the results presented in the main manuscript.

\subsection{Photoelectron distribution in a delay-line detector}
\begin{figure*}[ht]
	\includegraphics[width=1\textwidth]{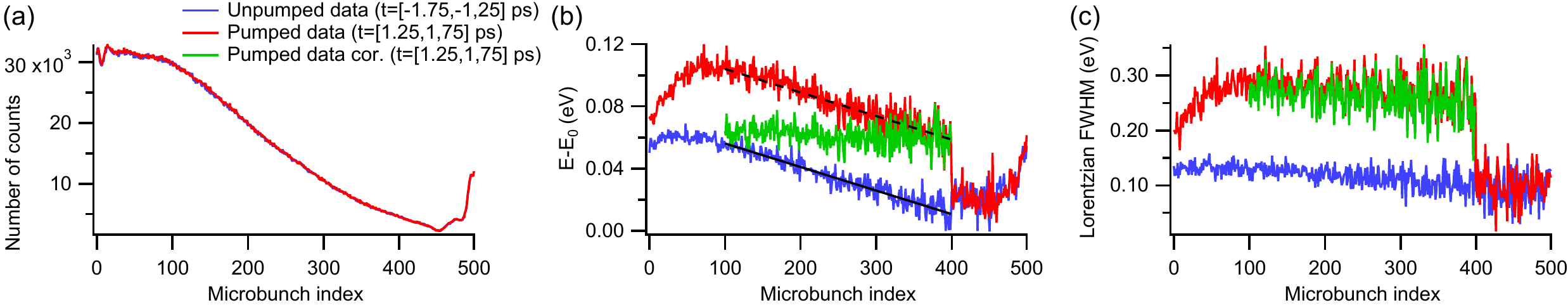}\\
	\caption{(a) Core-level peak position distribution over the four quadrant (Q1-Q4) DLD. Each pixel of the image corresponds to the fitted position of the W$4f_{7/2}$ core-level peak, represented by a false color scale. The blue line $y=x$ marks the line profile presented in the bottom panel. (b-c) The same data as in (a) after quadrant and spherical timing aberration corrections, respectively. See text for details.}
	\label{fig:fig5_sm}
\end{figure*}

The delay-line detector (DLD) used in this study consists of four independent modules (Q1-Q4), spanning the whole multichannel plate (MCP) surface, as presented in Fig.~S\ref{fig:fig2_sm}(a). In order to study the performance and relative timing of the modules, we examined the unpumped peak positions of the W $4f_{7/2}$ spectra as a function of the position on the detector, as illustrated in Fig.~S\ref{fig:fig5_sm}. Core-level spectra of heavy elements are perfect for this purpose due to their lack of dispersion. We find that the obtained peak position is nonuniform across the detection surface. This effect probably originates from $t_0$ timing differences of the individual DLD modules. Therefore, we apply a correction by adding a quadrant-dependent delay-time offset on the single-event level~\cite{Xian19}. This equalizes the distribution over the whole detection area, as shown in Fig.~S\ref{fig:fig5_sm}(b). Nevertheless, the whole constant-energy surface appears to be curved, which we attribute to the spherical timing aberration -- off-axis electrons travel longer distance than on-axis ones, artificially increasing the observed time-of-flight. Spherical timing aberration correction is described in detail elsewhere~\cite{Xian19}, and here we present just the final results (Fig.~S\ref{fig:fig5_sm}(c)). In addition, we filtered out events from the edges of the MCP as they were significantly off the mean value. After the artifact corrections, we find the standard deviation of the fitted peak positions over the whole detector to be ca. 10~meV.

\subsection{Effects of multi-hit events}
\begin{figure*}[ht]
	\includegraphics[width=1\textwidth]{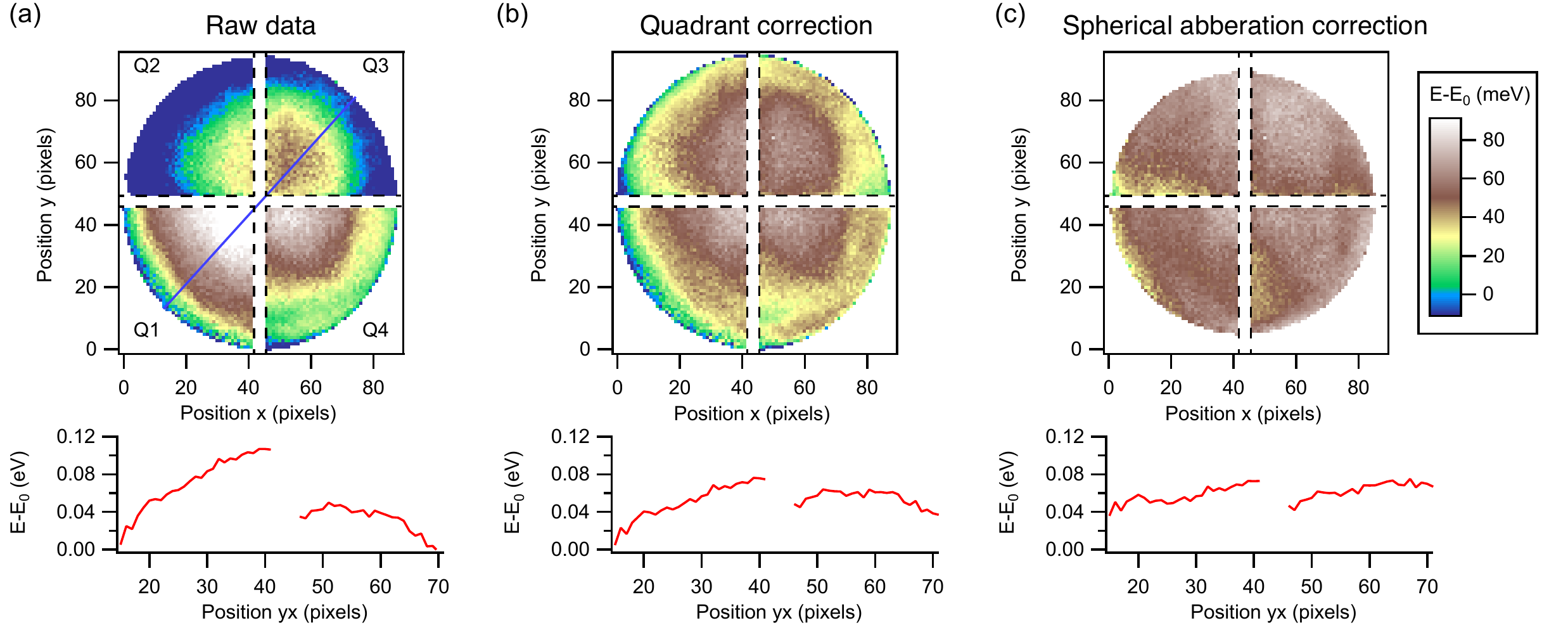}\\
	\caption{(a) Unpumped XPS spectra of W4f. Green, dashed lines marks the energy region of integration in (b). (b) Photoemission signal distribution over the whole DLD. Red, dashed circles mark four regions of increasing intensity. Green, dashed lines marks the y position region of integration in (c). (c) 2D histogram of the photoemission signal versus energy and position x on the detector. Cone-like increased intensity region, which correspond to DLD multi-hit artifact, is pointed to by an arrow. Blue (profile 1) and red (profile 2) lines mark spectra shown in (d-e), respectively. (f), Comparison of profiles 1 (multiplied by 1.28) and 2, which proves that the DLD artifact is not affecting the lineshape of the W$4f_{5/2}$ core-level peak.}
	\label{fig:fig6_sm}
\end{figure*}

DLDs are designed for single-hit detection and multi-hit events lead to experimental artifacts which confounds the photoemission signal. The detailed analysis of the unpumped W $4f$ XPS spectra revealed a small contribution for energies ca. 1~eV higher than W $4f_{7/2}$ (see Fig.~S\ref{fig:fig6_sm}(a)). This additional peak was not observed around W$4f_{5/2}$ peak, which indicate the artificial origin of this effect. The energy-integrated momentum-resolved 2D photoemission distribution on the DLD detectors shows four circular spots of higher intensity, approximately in the middle of every DLD quadrant (Fig.~S\ref{fig:fig6_sm}(b)). Due to the threefold symmetry of the sample material, we exclude the effect of photoelectron diffraction. Image integrated over one of the higher-intensity spots (Fig.~S\ref{fig:fig6_sm}(c)) clearly shows a cone-like feature starting from the W$4f_{7/2}$ peak. This feature is typically observed for DLD acquisition in the multi-hit regime and its influence on the observed core-level lineshape is investigated in Figs.~\ref{fig:fig6_sm}(d-e)-- Profile 1, close to the edge of a quadrant, is not affected, but Profile 2, through the center the quadrant, shows a small contribution close to the W$4f_{7/2}$ peak. Importantly, the lineshape of W$4f_{5/2}$ is not affected by the multi-hit artifact, as evidenced in Fig.~\ref{fig:fig6_sm}(f). Data analysis presented in the main text was conducted on W$4f_{5/2}$ spectra because of the above reason.

\subsection{Space-charge effects}
\begin{figure*}[ht]
	\includegraphics[width=1\textwidth]{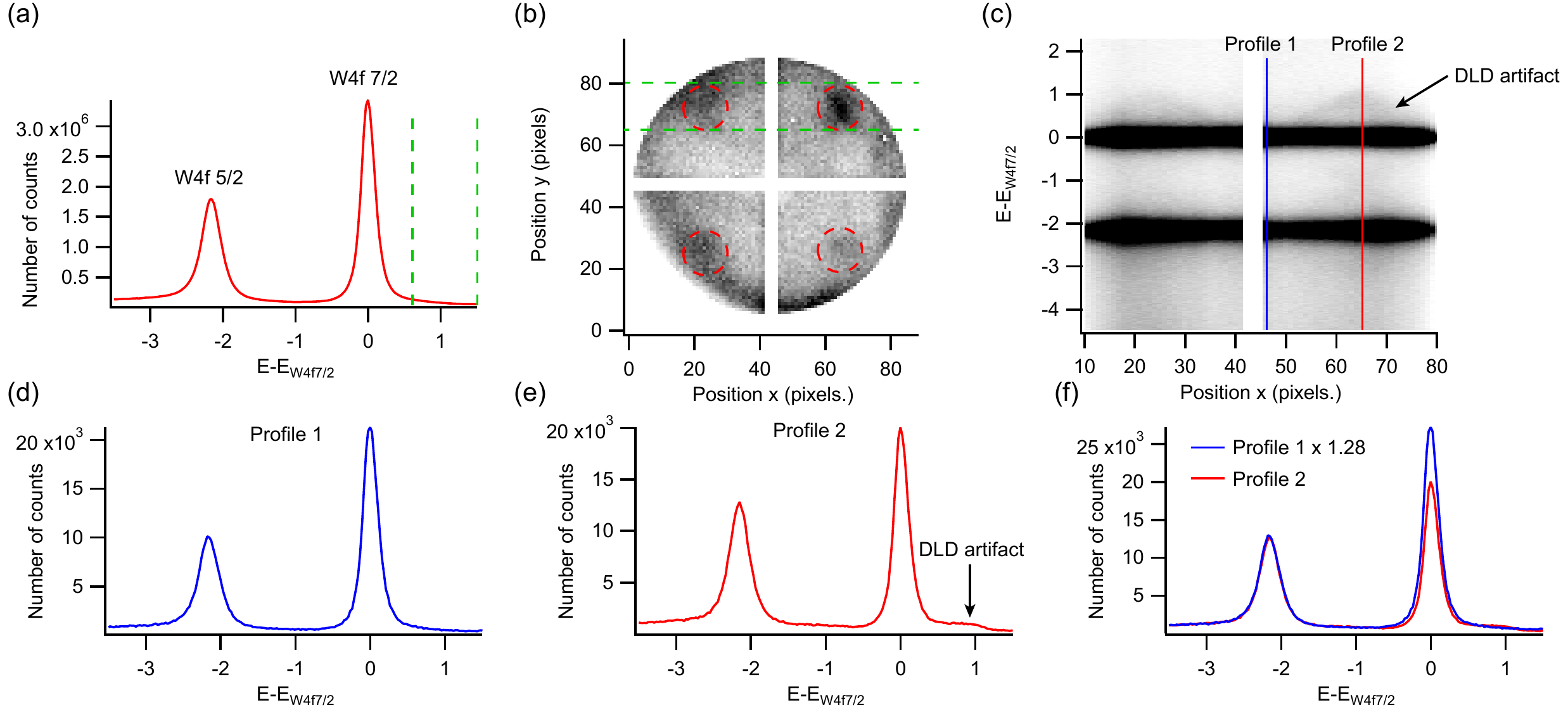}\\
	\caption{(a) Distribution of photoemission counts for corresponding microbunch index. (b-c) Fitted peak position and Lorentzian FWHM as a function of microbunch index, respectively. Black solid line represents a linear fit corresponding to microbunch-dependent probe-induced space charge. Black dashed line shows the same fit shifted by a constant. Blue, red and green lines correspond to unpumped, pumped and pumped, space-charge corrected data, respectively.}
	\label{fig:fig7_sm}
\end{figure*}

FELs produce intense and short x-ray pulses at relatively low repetition rate. This characteristic makes photoemission experiments challenging due to aforementioned multi-hit events and space-charge effects. In addition, we observed that the FEL pulses had a non-constant intensity distribution within a macrobunch, reflected in the variable photoemission yield presented in Fig.~S\ref{fig:fig7_sm}(a). We study the probe-induced space-charge effect by investigating the W$4f_{5/2}$ peak position and the FWHM (Figs.~\ref{fig:fig7_sm}(b-c)). For microbunches of indices in the range [100-399], which we use for trXPS measurements, the peak positions in both unpumped and pumped photoemission spectra shifts significantly following the number of counts. This behaviour is indicative of dominating effect of probe-induced space-charge and we correct it at the single-event level by adding a microbunch-dependent offset to the photoelectron energy. This correction flattens the peak position distribution, as shown in Fig.~S\ref{fig:fig7_sm}(b). The width of the W$4f_{5/2}$ peak is not affected significantly by the probe space-charge and no additional correction is needed (Fig.~S\ref{fig:fig7_sm}(c)). We note here that the applied energy correction is not affecting the width of peak.

\subsection{Pump-probe synchronization}
\begin{figure*}[ht]
	\includegraphics[width=1\textwidth]{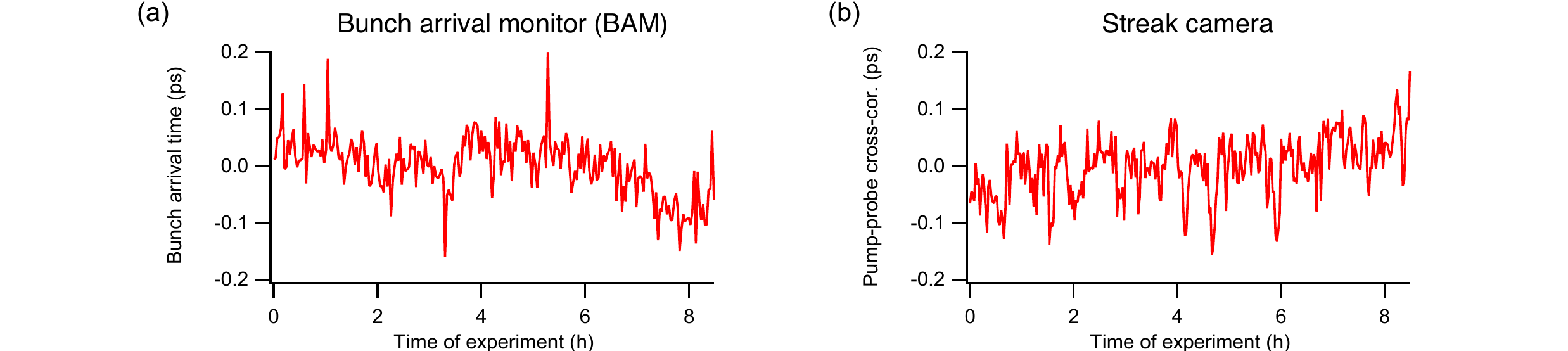}\\
	\caption{(a) Bunch arrival monitor (BAM) readings in respect to a common oscillator signal (every two-millionth point is plotted). (b) Streak camera readings measuring the pump-probe cross-correlation signal (every hundred-thousandth point is plotted).}
	\label{fig:fig8_sm}
\end{figure*}

Pump-probe experiments rely on a stable and well-defined time delay between pump and probe pulses. This is especially important for FELs operating in the self-amplified spontaneous emission (SASE) mode as FLASH. In this case, the temporal resolution is largely limited by the timing jitter~\cite{Ackermann07}. We correct for this on single-event level by subtracting the bunch arrival monitor (BAM)~\cite{Schulz15} readings from the delay-time. The long-term stability of the experimental setup is further monitored by a streak camera, measuring the pump-probe cross-correlation signal every ca. 60 pulses. Both BAM readings and the streak camera cross-correlation signals acquired during the presented experiment are shown in Figs.~\ref{fig:fig8_sm}(a-b), respectively. Overall, a stability on the order of 50~fs was maintained which resulted in a temporal resolution of ca. 160~fs.

\section{Low-fluence data}
\begin{figure*}[ht]
	\includegraphics[width=1\textwidth]{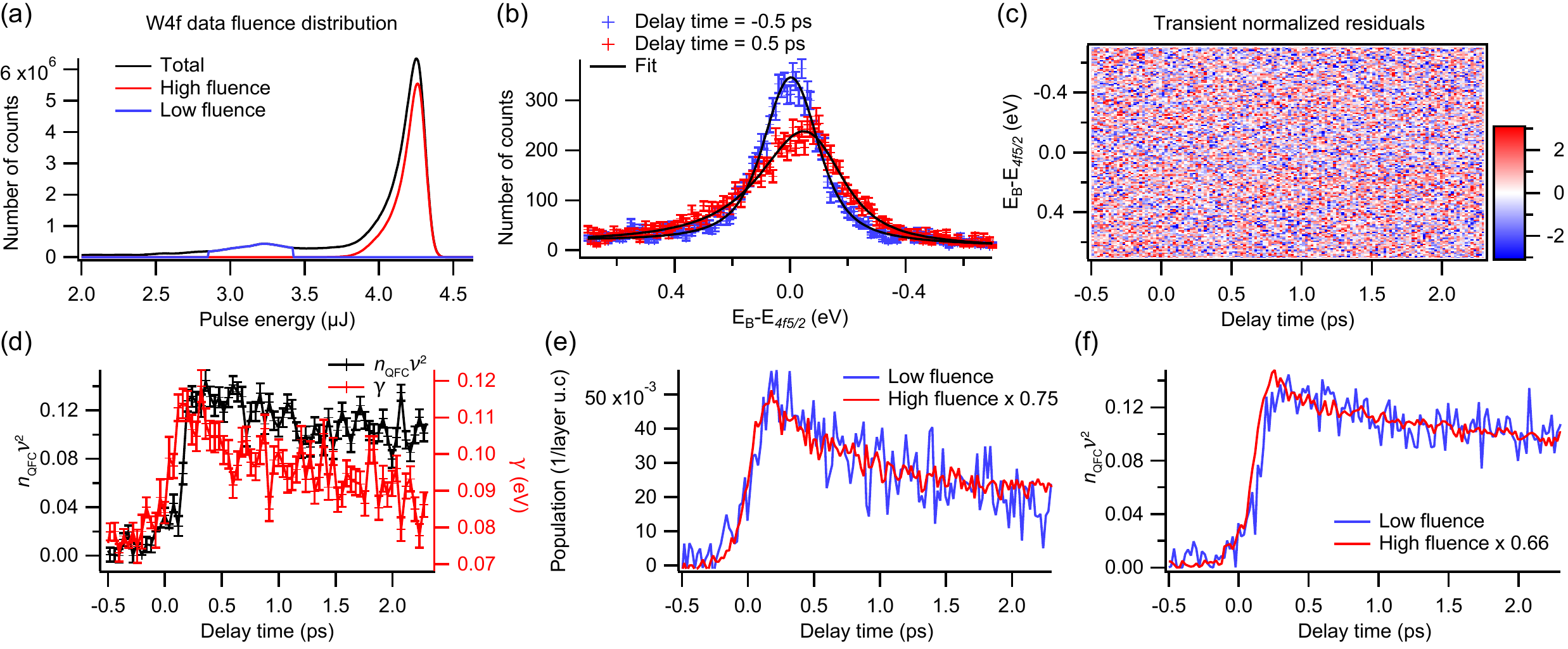}\\
	\caption{(a) trXPS W4f core-level fluence distribution. Black, red and blue lines correspond to total, high fluence and low fluence datasets, respectively. (b) Unpumped (blue), pumped (red) and fitted (black) W$4f_{5/2}$ spectra for the low fluence case. (c) Transient normalized residuals of the fitted spectral function over the whole analyzed range of pump-probe time delay. (d) Comparison of the time-dependent effective screening $n\nu^2$ (black) and broadening $\gamma$ (red) for the low fluence case. (e) Comparison of the excited population $n$ for the low fluence (blue) and high fluence (red, multiplied by 0.75) cases. (f) Comparison of the effective screening $n_{QFC}\nu^2$ for low fluence (blue) and high fluence (red, multiplied by 0.66) cases.}
	\label{fig:fig9_sm}
\end{figure*}

The excitonic Mott transition is a process driven by the excitation density, therefore, it is important to investigate this process at lower pump fluences. For this purpose, we utilize the photoemission data from the initial microbunches, as these were consistently pumped with lower fluence (see Fig.~S\ref{fig:fig4_sm}(a)) under exactly the same experimental conditions. Fig.~S\ref{fig:fig9_sm}(a) presents the pulse energy distribution for the whole dataset. For the high fluence data presented in the main text, the photoelectrons with microbunch indices in the range [100, 399] have been analyzed. This selection corresponds to the main peak in Fig.~S\ref{fig:fig9_sm}(a). We find, however, another peak of comparable shape in the pulse energy distribution, which correspond to 75\% of the high-fluence case. We use this subset of data to investigate the low-fluence case. We do not apply the microbunch-dependent space-charge correction as the low-fluence data consist of events with microbunch indices in the range [10, 55], for which probe-space charge is constant and affects only the reference W $4f$ peak position (see Fig.~S\ref{fig:fig7_sm}(b-c)). Although the low-fluence data consists of significantly fewer photoelectron events, it is still possible to obtain a reliable fit to the model spectral function presented in the main text (see Fig.~S\ref{fig:fig9_sm}(b-c). The obtained dynamical behaviour of the effective screening, $n_{QFC}\nu^2$, and the broadening, $\gamma$, shows qualitative agreement to the high-fluence case (see Fig.~S\ref{fig:fig9_sm}(d)), but we are able to identify a few important differences. Fig.~S\ref{fig:fig9_sm}(e) presents the low-high fluence comparison of the total population dynamics calculated from the corresponding $\gamma$ signals using the same calibration coefficients. We find an excellent match between these two cases after rescaling the high-fluence result by 75\%, which agrees well with independently-measured fluence difference and corroborates our expectations concerning the data interpretation discussed in the main text. Furthermore, the dynamics of $n_{QFC}\nu^2$ after 0.5~ps also show a similar decay, but only after rescaling the high-fluence data by 66\%. Based on these two observations, we conclude that in the low-fluence case, the Mott transition is not complete and ca. 9\% of the excited carriers stay in the excitonic state and thus do not contribute to the effective screening of the core hole. Additionally, the dynamics of $n_{QFC}\nu^2$ at low fluences seem to be slower than in the high-fluence case within the initial 0.5~ps time delay range, indicating that the excitonic Mott transition rate is dependent on the excitation density.
\section*{Supplementary references}
\bibliographystyle{apsrev}
\bibliography{references_SM}
\bibliographystyle{apsrev}